%% file: ms.tex
\documentclass[twocolappendix]{emulateapj}
\usepackage{amsmath}
\usepackage{subfigure}
\usepackage{graphicx}
\usepackage{color}

\newcommand{\tr}{\tilde{r}}

\newcommand{\V}[1]{ \mathbf #1 }
\newcommand{\Alfven}{Alfv{\'e}n }

\newcommand{\figext}[1]{#1.pdf}

\newcommand{\Section}[1]{Section~\ref{sec:#1}}
\newcommand{\Appendix}[1]{Appendix~\ref{sec:#1}}
\newcommand{\Equation}[1]{Equation~\ref{eqn:#1}}
\newcommand{\Figure}[1]{Figure~\ref{fig:#1}}
\newcommand{\Table}[1]{Table~\ref{tab:#1}}

\include{figures}

\begin{document}

\title{Freely decaying turbulence in force-free electrodynamics}

\author{Jonathan Zrake and William E.\ East}

\affil{Kavli Institute for Particle Astrophysics and Cosmology, Stanford
  University, SLAC National Accelerator Laboratory, Menlo Park, CA 94025, USA}

\keywords {
  magnetohydrodynamics ---
  turbulence ---
  magnetic fields ---
  gamma-rays: bursts ---
}

\begin{abstract}
  Freely decaying relativistic force-free turbulence is studied for the first
  time. We initiate the magnetic field at a short wavelength and simulate its
  relaxation toward equilibrium on two and three dimensional periodic domains,
  in both helical and non-helical settings. Force-free turbulent relaxation is
  found to exhibit an inverse cascade in all settings, and in 3D to have a
  magnetic energy spectrum consistent with the Kolmogorov $5/3$ power law. 3D
  relaxations also obey the Taylor hypothesis; they settle promptly into the
  lowest energy configuration allowed by conservation of the total magnetic
  helicity. But in 2D, the relaxed state is a force-free equilibrium whose
  energy greatly exceeds the Taylor minimum, and which contains persistent
  force-free current layers and isolated flux tubes. We explain this behavior in
  terms of additional topological invariants that exist only in two dimensions,
  namely the helicity enclosed within each level surface of the magnetic
  potential function. The speed and completeness of turbulent magnetic free
  energy discharge could help account for rapidly variable gamma-ray emission
  from the Crab Nebula, gamma-ray bursts, blazars, and radio galaxies.
\end{abstract}

\maketitle

\section{Introduction} \label{sec:introduction}

The most extreme sources of high energy astrophysical radiation are widely
believed to exist in magnetically dominated, relativistic environments. Jets
powered by super-massive black holes, plasma winds driven by pulsars, and
gamma-ray bursts are prime examples. The violent intermittency of gamma-ray
production by these systems could be taken as strong evidence that turbulence is
critically linked to their radiative output. And yet, the physics of
magnetically dominated, relativistic turbulence remains nearly unexplored.

The importance of understanding turbulence in this new regime is underscored by
the discovery of powerful gamma-ray flares originating within the Crab Nebula
\citep{Abdo2011, Tavani2011}. Moreover, rapid time variability seems to be
ubiquitous among gamma-ray emitters; the blazars PKS 2155-304
\citep{Aharonian2007}, 1510-089 \citep{Saito2013}, and 3C 279
\citep{Hayashida2015}, as well as radio galaxies such as M87
\citep{Aharonian2006} and IC 310 \citep{Aleksic2014}, have each been observed to
produce sporadic, high intensity outbursts of gamma-radiation. Such dramatic
enhancements of synchrotron or inverse Compton emissivity require a reservoir of
free energy to spontaneously energize the active region's electron
population. If that free energy resides in magnetic fields, then its discharge
could be triggered by magnetic reconnection --- the general picture of which has
been rendered in many different ways \citep{Lyutikov2003a, Lazarian2003,
  Zhang2011, McKinney2012a, Sironi2014, Blandford2015}.

In this paper we intend to demonstrate that magnetic free energy discharge can
proceed from field geometries that are far more general than those typically
considered in reconnection models, and on a time scale that is not limited by
the rate with which microphysical or anomalous \citep[e.g.][]{Lazarian1999}
resistivity can destroy magnetic flux. This amounts to extending the historical
problem of magnetic relaxation \cite[e.g.][]{Chandrasekhar1953} to relativistic,
magnetically dominated conditions. We focus on only a few of the many aspects of
this topic that could be studied.
Briefly, they are: (1) the rate and
completeness of magnetic free energy discharge in various topological settings,
(2) a characterization of persistent non-linear structures, and (3) the spectral
energy distribution of freely decaying relativistic force-free turbulence. To be
most relevant for astrophysical gamma-ray emission, we are interested in regions
far from any solid boundaries that could anchor the magnetic field (so periodic
domains are appropriate), and where the plasma is nearly perfectly conducting,
inviscid, and magnetically dominated --- conditions which are the domain of
force-free electrodynamics (FFE) theory.

Force-free electrodynamics forms the basis for historical theories of pulsar
magnetospheres \citep{Goldreich1969, Spitkovsky2006} and angular momentum
extraction from black holes \citep{Blandford1977}, and continues to be a widely
used description for studying these highly relativistic settings
\citep{Palenzuela2010, Yang2015, Gralla2015}. It can be derived from
relativistic magnetohydrodynamics (MHD) when the electromagnetic contribution to
the stress-energy tensor greatly exceeds contributions from matter, and hence it
captures the essential non-linear dynamics of relativistic MHD for the regime of
interest. It also admits a numerical approach that is more robust and efficient
than relativistic MHD solution schemes.

Turbulence in force-free electrodynamics has only been considered in a few
previous studies. The theory of \Alfven wave turbulence in the presence of a
strong guide field, originally formulated for Newtonian MHD by
\cite{Goldreich1995}, has been extended to the magnetically dominated,
relativistic regime by \cite{Thompson1998}. \Alfven wave turbulence has since
been studied numerically in both the momentum balanced \citep{Cho2005} and
unbalanced \citep{Cho2014} situations. Even the study of mildly relativistic MHD
turbulence is in its infancy, having only been treated so far in a handful of
studies \citep{Zhang2009a, Inoue2010, Zrake2011, Zrake2013, Zrake2014}. There
are, by comparison, a great number of Newtonian MHD turbulence studies \cite[see
  e.g.][for a review]{Tobias2011} treating all different circumstances,
including turbulent relaxation. Comparisons with them will be made wherever
possible.

One of the issues we will explore in this paper is the applicability of the
\cite{Taylor1974} hypothesis to magnetic relaxation in force-free
electrodynamics. Taylor's original conjecture was that magnetic relaxation would
universally settle in the lowest energy configuration allowed by the
conservation of total magnetic helicity
\begin{equation} \label{eqn:helicity}
  H = \int \V A \cdot \V B d^3x.
\end{equation}
These so-called Taylor states are \emph{linear} force-free equilibria, having
electric current density $\V J$ that is not only aligned with the magnetic
field, but is also uniformly proportional to it, i.e. they solve the constraint
\begin{equation} \label{eqn:force-free-condition}
  \nabla \times \V B = \alpha \V B
\end{equation}
for a global inverse length scale $\alpha$. Such field configurations are
\emph{monochromatic}, all their magnetic energy is concentrated around the
spatial frequency $\alpha$. The converse of Taylor's conjecture is that
relaxation may, in some circumstances, end in a more general force-free
equilibrium in which $\alpha$ could vary from one magnetic field line to
another. In such non-linear equilibria, the highest values of $\alpha$ are
associated with the smallest scale coherent structures, which may be current
layers or flux tubes, and are associated with peaks in the intensity of
electrical current flow.

Counterexamples to Taylor's conjecture do exist, but those identified so far
apply to settings in which gas pressure plays a role. For example, hydromagnetic
relaxed states with non-uniform $\alpha$ were reported by \cite{Amari2000} and
\cite{Pontin2013} where the magnetic field lines terminate on conducting plates,
a boundary condition that is motivated by the physics of the solar corona. More
general hydromagnetic equilibria have also been found in simulations of
stratified environments such as stellar interiors, a setting that has been
extensively explored by \cite{Braithwaite2006, Braithwaite2008, Braithwaite2009}
and \cite{Duez2010a}. \cite{Gruzinov2009} followed incompressible MHD relaxation
of a non-helical magnetic field in two dimensions \footnote{Gruzinov's
  two-dimensional simulations followed only the in-plane magnetic field. In the
  rest of this paper, ``two-dimensional'' means that translational symmetry is
  enforced along the $z$-axis, but $B_z$ need not vanish. This setting is
  sometimes referred to as 2.5D.} and found that it did not decay toward the
Taylor minimum (total annihilation of the field in this case), but instead was
halted in an approximate equilibrium with many current layers, beyond which
further decay was only made possible by slow resistive evolution.

Our study makes frequent use of the periodic short-wavelength Taylor states as
initial conditions. A Taylor state of frequency $\alpha_0$ and helicity $H$ has
an energy $\alpha_0 H / 2$, a fraction $1 - \alpha_1/\alpha_0$ of which could be
dissipated without changing the total helicity (where $\alpha_1 = 2\pi/L$ is the
lowest allowed frequency, although we will use $L = 2\pi$ so that $\alpha_1 =
1$). This implies that their free energy supply can be arbitrarily large, and so
raises the question of their mechanical stability. Very recently,
\cite{East2015} found that in FFE as well as in relativistic MHD, generic
examples of the 3D, periodic $\alpha_0 > 1$ Taylor states are unstable to small,
ideal perturbations, with a growth rate that is proportional to the inverse
\Alfven time. Upon saturation of the linear instability, decay enters a
turbulent stage that lasts until the remaining energy $\alpha_1 H / 2$ resides
at the lowest allowed frequency $\alpha_1$. This behavior bears out the
predictions of \cite{Frisch1975} which were based on the prediction that
turbulence would generically shift magnetic helicity toward large scales.

Conventionally, this so-called inverse cascade has been thought to operate
efficiently only when the field is strongly helical, a belief which has dramatic
consequences for large-scale dynamo theory \citep{Blackman2004}, as well as the
evolution of cosmic magnetic fields since the early universe \citep{Olesen1997,
  Son1999, Banerjee2004}. However, an efficient inverse cascade was recently
observed to occur even when the field was fully non-helical, in both Newtonian
\citep{Brandenburg2015} and relativistic \citep{Zrake2014} MHD
settings. Although the magnetic energy eventually decays toward zero, the
relaxation evolves in a self-similar manner, depositing energy in structures
larger than the coherence scale $k_B^{-1}$, which increases over time until it
attains the system size. In this study we will show that all settings of freely
decaying turbulence in force-free electrodynamics, 2D and 3D, helical and
non-helical, exhibit inverse cascading. The 2D and helical case is particularly
fast and nearly conservative; as time goes on, magnetic energy is shifted toward
ever-increasing scales while suffering a diminishing rate of dissipative losses.

Our paper is organized as follows. We briefly describe the theory of force-free
electrodynamics and its invariants in \Section{FFE}. There we also discuss the
special case of two dimensions, and define the additional topological invariants
that it imposes. We then outline the numerical scheme that is used to solve the
FFE equations in \Section{methods}, and describe our numerical implementation of
various diagnostics such as power spectra, characteristic scales, and the
helicity invariants. Our simulation results, including the energy of relaxed
magnetic configurations, an analysis of coherent structures, spectral energy
distributions, and details of the inverse cascade, are presented in
\Section{results}. We discuss the implications of these results for
astrophysical gamma-ray sources in \Section{discussion}, and also point out how
our results might aid in the interpretation of two-dimensional (including
axisymmetric) calculations. \Appendix{numerical-convergence} contains some
details on the numerical convergence of our scheme. Throughout our paper, we use
units in which the speed of light $c=1$. The domain scale $L$ is set to $2 \pi$
so that the smallest spatial frequency is 1, and time is reported in units of
the light-crossing time $L / c$.

\section{Force-free electrodynamics and its invariants}
\label{sec:FFE}

\begin{deluxetable}{cccccc}
  \tablecolumns{4}
  \small
  \tablewidth{0pt}
  \tablecaption{Summary of runs}

  \tablehead{
    \colhead{Grid resolution} &
    \colhead{$\epsilon$} &
    \colhead{$\alpha_0$} &
    \colhead{See Figure}}
  
  \tablecomments{Shown is a summary of the runs along with the figures they are
    referenced by. Helical runs have $\epsilon=1$ and non-helical runs have
    $\epsilon=0$. Those whose initial data is the 2D ABC field have $\alpha_0$
    values that are exact integers, whereas randomized initial data have
    $\alpha_0$ values that are not exact integers.}

  \startdata

  $512^3$  & 1 & $16$  & \ref{fig:a256-H1-3072-imgs}b \nl
  $512^2$  & 1 & $16$ & \ref{fig:energy-decay-time-series} \nl
  $512^3$  & 1 & $\sqrt{257} \approx 16$ & \ref{fig:energy-decay-time-series} \nl
  $512^2$  & 0 & $\sqrt{257} \approx 16$ & \ref{fig:energy-decay-time-series} \nl

  $3072^2$ & 1 & $256$ & \ref{fig:a256-H1-3072-imgs}a,
  \ref{fig:a256-H1-3072-prof1d}, \ref{fig:a256-H1-alpha-hist},
  \ref{fig:magnetic-bubble-profile} \nl

  $4096^2,6144^2,8192^2,12288^2$ & 1 & $256$ &
  \ref{fig:a256-H1-tseries-convergence} \nl

  $16384^2$ & 1 & $256$ & \ref{fig:a256-H1-16384-Pbfit},
  \ref{fig:a256-H1-tseries-convergence} \nl

  $256^2,512^2,1024^2,2048^2,4096^2$ & 1 & $8,16,32,64,128$ &
  \ref{fig:alpha-hist-compare} \nl

  $4096^2$ & 1 & $32$ & \ref{fig:helicity-dist} \nl
  
  $12288^2$ & 0 & $\sqrt{65641} \approx 256$ &
  \ref{fig:energy-decay-time-series}, \ref{fig:pspec-compare},
  \ref{fig:scale-evolution}a \nl
  $12288^2$ & 1 & $\sqrt{65641} \approx 256$ &
  \ref{fig:energy-decay-time-series}, \ref{fig:pspec-compare},
  \ref{fig:scale-evolution}b \nl
  
  $1024^3$  & 0 & $\sqrt{2305} \approx 48$ & \ref{fig:pspec-compare},
  \ref{fig:scale-evolution}c \nl
  $1024^3$  & 1 & $\sqrt{2305} \approx 48$ & \ref{fig:pspec-compare},
  \ref{fig:scale-evolution}d \nl

  $16384^2$ & 0 & $\sqrt{65641} \approx 256$ & \ref{fig:pspec-compare} \nl 
  $16384^2$ & 1 & $\sqrt{65641} \approx 256$ & \ref{fig:pspec-compare} \nl
  $768^3$   & 0 & $\sqrt{2305} \approx 48$ & \ref{fig:pspec-compare} \nl
  $768^3$   & 1 & $\sqrt{2305} \approx 48$ & \ref{fig:pspec-compare} \nl
  
  \enddata
  \label{tab:run-summary}
\end{deluxetable}

Force-free electrodynamics describes the flow of electromagnetic energy in a
charge-supplied medium with vanishing inertia. This approximation is useful for
plasma environments in which the energy density of the magnetic field greatly
exceeds contributions from the matter. Under such conditions, the flow of
electrical current responds rapidly to changes in $\V E$ and $\V B$ in order to
cancel the Lorentz force density $\rho \V E + \V J \times \V B$, where $\rho =
\nabla \cdot \V E$ is the net electrical charge per unit volume. FFE thus admits
an Ohm's law that is a function of $\V E$ and $\V B$ alone,
\citep[e.g.][]{McKinney2006a}
\begin{equation} \label{eqn:ffe-ohm}
  \V J = \frac{\V B}{B^2}\left( \V B \cdot \nabla \times \V B - \V E \cdot
  \nabla \times \V E \right) + \frac{\V E \times \V B}{B^2} \rho,
\end{equation}
which may be coupled to the Maxwell equations
\begin{eqnarray} \label{eqn:maxwell}
  \partial_t \V E &=& \phantom{-} \nabla \times \V B - \V J \\
  \partial_t \V B &=&          -  \nabla \times \V E \nonumber
\end{eqnarray}
in order to yield a hyperbolic system of partial differential equations
governing the evolution of the six components of the electromagnetic field
\citep{Pfeiffer2013}.

Evolution of ideal MHD systems in general is constrained by three quadratic
invariants --- the total energy $U$, the magnetic helicity $H$, and the cross
helicity $W = \int \V v \cdot \V B d^3x$ where $\V v$ is the fluid velocity
\citep{Bekenstein1987}. As a limiting case of MHD, force-free electrodynamics
shares these invariants, with the exception of the cross helicity, since FFE
does not define the fluid velocity in the direction of $\V B$. Most relevant to
our study of magnetic relaxation is conservation of magnetic helicity, which is
a topological invariant of the magnetic field alone, and thus functions in the
same way for MHD as it does for FFE. In both MHD and FFE, $H$ is a robust
invariant, as it is generally found to be conserved even in the presence of
small non-ideal effects \cite[see e.g.][]{Blackman2014}.

\subsection{Energy}

\figureImageSequence

Since the Lorentz force density vanishes in FFE, no $\V E \cdot \V J$ work is
done on the charge carriers and the system is formally energy conserving.
Nevertheless, time-dependent solutions may develop regions in which the
condition $E < B$ is violated, i.e. no frame exists in which the electric field
vanishes. This indicates a breakdown of the ideal force-free assumption, and the
Ohm's law given by \Equation{ffe-ohm} must be modified. Although non-ideal
force-free Ohm's laws have been proposed in the literature \citep{Gruzinov2007,
Li2012}, it is still commonplace to evolve the ideal system until such a
breakdown occurs, and when it does to simply reduce the magnitude of $\V E$ to
prevent $E > B$.  The physical motivation for this prescription, which we use
here, is that energy is being radiated away when charges are accelerated to
short out the electric field and restore $ E < B$. The numerical evolution
scheme is described in more detail in \Section{numerical-scheme}, and in
\Appendix{numerical-convergence} we confirm that it yields numerical convergence
of the energy dissipation rate.

\subsection{Topology} \label{sec:topology}

Force-free electrodynamics shares the same magnetic topological invariants as
Newtonian and relativistic MHD, the lowest order of which is the total magnetic
helicity given by \Equation{helicity}. Its invariance can be seen as stemming
from the conservation of magnetic flux through a closed field loop, which is why
it is commonly referred to as characterizing the interlocking of the magnetic
field. Although the helicity density $\V A \cdot \V B$ depends on the choice of
gauge, its integral $H_{\mathcal{V}}$ over any volume $\mathcal{V}$ bounded by
magnetic surfaces is well-defined, and also invariant under ideal evolutions,
such as Faraday's law, when $\V E \cdot \V B = 0$
\citep[e.g.][]{Brandenburg2005},
\begin{equation} \label{eqn:helicity-change}
  \partial_t H_{\mathcal{V}} = -2 \int_{\mathcal{V}} \V E \cdot \V B d^3x.
\end{equation}
\Equation{helicity-change} implies that $H_{\mathcal{V}}$ can still be conserved
approximately in the presence of non-ideal processes, as long as the volume in
which they occur is small.

In principle, a domain that admits a partitioning by magnetic surfaces has an
independently conserved helicity associated with each smaller volume. But in
practice, three dimensional field geometries are too complex for such a
partitioning to be possible. This is what led Taylor to conclude that relaxation
is only constrained by a single topological invariant --- the total
helicity. The story is different in two dimensions since the magnetic surfaces
are far simpler; their cross-sections are nested closed curves in the $x-y$
plane, and the helicity enclosed by each functions as an independently conserved
quantity for as long as that surface retains its identity. But non-ideal
effects, however small, permit the surfaces to merge with one another, erasing
their identities and shuffling up their conserved helicities. Nevertheless, in
two dimensions we can still construct a robust topological invariant that is far
stricter than the total helicity.

We recall that in the presence of $z$-translational symmetry, the in-plane
magnetic field is tangent to the isocontours of the magnetic flux function
$A_z$. So each sub-volume $\mathcal{V}_i(\psi)$ in which $A_z > \psi$ is
associated with a conserved helicity $\mathcal{H}_i(\psi) =
\int_{\mathcal{V}_i(\psi)} \V A \cdot \V B d^3x$. To whatever extent the
helicities of such volumes remain additive with respect to reconnections between
their bounding surfaces, the sum $\mathcal{H}(\psi) = \sum_i
\mathcal{H}_i(\psi)$ must be a robust topological invariant. This assumption is
justified by \Equation{helicity-change} when the non-ideal regions (where $\V E
\cdot \V B \ne 0$) are small. Alternatively, consider the helicity
\citep[e.g.][]{Berger1984}
\begin{equation}
  \mathcal{H}_{12} = 2 \Phi_1 \Phi_2
\end{equation}
of two flux tubes that are once-linked and independently non-helical. An example
is a cylindrical structure whose external, toroidal flux $\Phi_2$ wraps an
interior, poloidal flux $\Phi_1$. Two such structures that were joined to share
an outer surface would then enclose a poloidal flux $2 \Phi_1$, which is wrapped
by the same toroidal flux $\Phi_2$. So, to whatever extent the ``joining'' is
done without destroying magnetic flux, the resulting arrangement has a helicity
$2 \mathcal{H}_{12}$.

The presence of additional invariants $\mathcal{H}(\psi)$ is expected to place a
more restrictive lower bound on the magnetic energy than would the total
helicity $H$ alone. In particular, given that Taylor states of the same helicity
but different wavelength will generally span a distinct range of $\psi$ values,
there may be no way for one 2D linear equilibrium to evolve into another of a
different wavelength while leaving $\mathcal{H}(\psi)$ unchanged. With this in
mind, we anticipate that fully relaxed 2D states will have an energy that
exceeds the Taylor minimum of $\alpha_0 H / 2$. We will describe the numerical
determination of $\mathcal{H}(\psi)$ in \Section{diagnostics}, and
in \Section{hel-distrib} we confirm that it is conserved by directly measuring
it in our 2D simulations.


There is also the question of what should happen to configurations in which $H =
0$ but $\mathcal{H}(\psi) \ne 0$, since such a configuration could not attain
arbitrarily low energy while respecting each of the invariants
$\mathcal{H}(\psi)$. Although behavior like this may be entirely possible, the
particular initial conditions used in this study (described in
\Section{initial-conditions} and summarized in \Table{run-summary})
generally have values of $|\mathcal{H}(\psi)|$ that are much smaller than $2 U_B
/ \alpha_1$, so we have not been able to observe it yet. Indeed, we will see
in \Section{energy-relaxed} that our 2D states with zero net helicity still
decay toward very small energy.

\section{Methods} \label{sec:methods}

We simulate magnetically dominated, relativistic turbulence on a periodic
domain of length $L = 2\pi$ in either two or three dimensions, using solutions
of the ideal force-free electrodynamics equations, given by \Equation{ffe-ohm}
and \Equation{maxwell}.

\subsection{Numerical scheme} \label{sec:numerical-scheme}

We evolve the FFE system using a fourth-order finite difference scheme. We use
standard fourth-order difference operators to evaluate all the field
gradients, and standard fourth-order Runge-Kutta time-stepping to advance the
solution in time.

The FFE system requires three vector constraints to be maintained: no monopoles
$\nabla \cdot \V B = 0$, perfect conductivity $\V E \cdot \V B = 0$, and the
existence of a frame in which $\V E$ vanishes $E < B$. The first two constraints
are formally preserved by FFE, but can be violated numerically at the level of
truncation error. Our scheme maintains the solenoidal constraint using the
hyperbolic divergence cleaning scheme proposed by \cite{Dedner2002}, and later
used in FFE simulations \citep[e.g.][]{Palenzuela2010}. This amounts to
supplementing Faraday's law with a magnetic monopole current $\V J_B = -\nabla
\Psi$, where the scalar field $\Psi$ evolves according to the damped wave
equation $\partial_t \Psi = - \nabla \cdot \V B - \tau^{-1} \Psi$ with $\tau$
being a non-physical time scale for quenching the magnetic monopole. $\V E \cdot
\V B = 0$ is maintained exactly by disregarding the part of the truncation error
that would give rise to a component of $\V E$ in the direction of $\V
B$. Numerical noise introduced by finite difference operations can lead to
unphysical growth of modes whose wavelengths are comparable to the numerical
grid spacing. Our scheme suppresses these unphysical modes using Kreiss-Oliger
dissipation, a form of low-pass filtering.  Each of the procedures just
mentioned supplements the FFE equations with terms at or below the level of the
truncation error, so they do not modify the formal convergence order of our
numerical scheme.

This numerical scheme was used in \cite{East2015}, and convergence results, as
well as comparisons to relativistic MHD simulations and analytical methods can
be found in that reference. It has been implemented as part of the \texttt{Mara}
\citep{Zrake2011} suite of relativistic turbulence codes, which has many
run-time post-processing capabilities that allow us to perform spectral and
statistical analysis of the solution at a high cadence while minimizing strain
on the host architecture's filesystem.

\subsection{Initial conditions} \label{sec:initial-conditions}

We start our simulations with a monochromatic magnetic field,
where all the power is at a single wavenumber magnitude $\alpha_0$, 
and with a vanishing electric field. The
general expression for our initial conditions is
\begin{eqnarray} \label{eqn:general-ic}
  \V B(\V x) &=& \sum_{|\V k|=\alpha_0} (\alpha_0 \V \Psi_{\V k} + \epsilon i \V
  k \times \V \Psi_{\V k}) e^{i \V k \cdot \V x} \\ \V k \cdot \hat{\V \Psi}_{\V
    k} &=& 0 \nonumber \\ \hat{\V \Psi}_{\V{k}} &=& \hat{\V \Psi}_{-\V{k}}^{*}
  \nonumber
\end{eqnarray}
where the parameter $\epsilon$ is chosen to be either one or zero,
corresponding to helical or non-helical configurations, respectively. Helical
initial configurations where $\alpha_0 > 1$ are unstable equilibria (see
\Section{linear-instability}), whereas the non-helical configurations
are out of equilibrium. Some of our initial conditions are randomized, having
$\V \Psi_{\V k} = \hat{\V e}_{\V k} e^{i \phi_{\V k}}$ where $\hat{\V e}_{\V k}$
is a random unit vector in the plane orthogonal to $\V k$ and $\phi_{\V k}$ is a
random phase. We also make use of a special case of \Equation{general-ic} known
as the ``ABC'' solution \citep{Arnold1965, Dombre1986}. In general this is given
by
\begin{equation} \label{eqn:ABC}
  \V B^{\rm{ABC}}(\V x) = \begin{pmatrix} B_3\cos \alpha_0 z - B_2\sin \alpha_0
    y \\ B_1\cos \alpha_0 x - B_3\sin \alpha_0 z \\ B_2\cos \alpha_0 y - B_1\sin
    \alpha_0 x \\
  \end{pmatrix},
\end{equation}
which is highly ordered and fully helical, meaning that $\V B = \alpha_0 \V A$
(in the Coulomb gauge). In this study we will make frequent use of the case with
$B_1 = B_2 = 1$ and $B_3 = 0$, which we refer to as the 2D ABC configuration.

Our results are based on simulations having a range of initial frequencies
$\alpha_0$ and numerical resolutions --- which we will refer to by the number of
grid points in each linear dimension $N$.  In general, the quality of our
results improves when we are able to simulate larger values of $\alpha_0$ with
more separation between the initial length scale and the domain length scale.
However, features (of size $\alpha_0^{-1}$) in our initial condition need to be
resolved by a certain number of grid points in order to obtain robust
solutions. In \Appendix{numerical-convergence} we show that 32 cells per
$\alpha_0^{-1}$ are sufficient to keep the error in the global helicity
conservation smaller than 1\%. In 2D we will present simulations with $\alpha_0$
as large as 256, with resolutions up to $16384^2$. In 3D, we will present
simulations with $\alpha_0$ as large as 48 and resolution $1024^3$.

\subsection{Diagnostics} \label{sec:diagnostics}

We define the power spectral density of the electric, magnetic, and helicity
fields, respectively, as
\begin{eqnarray} \label{eqn:power-spectra}
  P_E(k_i) &=& \frac{1}{\Delta k_i}\sum_{k_i < |\V q| < k_i + \Delta k_i} \V
  E_{\V q} \cdot \V E^*_{\V q} / 2, \\ P_B(k_i) &=& \frac{1}{\Delta k_i}\sum_{k_i
    < |\V q| < k_i + \Delta k_i} \V B_{\V q} \cdot \V B^*_{\V q} / 2 , \nonumber
  \\ P_H(k_i) &=& \frac{1}{\Delta k_i}\sum_{k_i < |\V q| < k_i + \Delta k_i} \V
  A_{\V q} \cdot \V B^*_{\V q} \nonumber
\end{eqnarray}
where $\V E_{\V q}$, $\V B_{\V q}$, and $\V A_{\V q}$ are, respectively, the electric field,
magnetic field, and vector potential Fourier harmonics of wavenumber $\V q$. We 
normalize the Fourier harmonics so that the volume integrated electric and
magnetic field energies $U_E$ and $U_B$, and the magnetic helicity $H$ are given
by
\begin{eqnarray*}
  U_E = \sum_{i} P_E(k_i) \Delta k_i , \\
  U_B = \sum_{i} P_B(k_i) \Delta k_i , \\
  H   = \sum_{i} P_H(k_i) \Delta k_i .
\end{eqnarray*}
We also define the characteristic frequency of each field $k_E$, $k_B$, and
$k_H$ as
\begin{equation} \label{eqn:k_X}
  k_X = \frac{\sum_i{P_X(k_i) k_i \Delta k_i}}{\sum_i{P_X(k_i)} \Delta k_i}
\end{equation}
where $X$ is one of $E$, $B$, or $H$. The most probable wavenumber, where 
$P_X(k)$ is maximal, is denoted by $\tilde{k}_X$.

In two dimensions, we track the ``helicity mass'' function discussed
in \Section{topology},
\begin{equation} \label{eqn:helicity-psi}
  \mathcal{H}(\psi) = \int \Theta (A_z(\V x) -\psi) \V A \cdot \V B d^3x,
\end{equation}
where $\Theta$ is the Heavyside step function. In practice, this diagnostic is
more easily computed as the ``helicity density'' function $d\mathcal{H}/d\psi$,
which we calculate by binning the lattice points according to their value of
$A_z$, and assigning the weight $\V A \cdot \V B$. We also create the volume
distribution $d\mathcal{V}/d\psi$ by binning points according to $A_z$ with
uniform weights, and the helicity distribution over volume
$d\mathcal{H}/d\mathcal{V} = \frac{d\mathcal{H}}{d\psi} /
\frac{d\mathcal{V}}{d\psi}$.

\section{Results}\label{sec:results}

\Figure{a256-H1-3072-imgs} shows the evolution of both two and three dimensional
freely decaying force-free magnetic turbulence. Both of these calculations are
initiated in the 2D ABC state, but the one on top takes place on a
two-dimensional domain where translational symmetry is assumed in the $z$
direction, and the bottom one was given a low-level white-noise perturbation to
break the $z$-symmetry. The left-most image shows the solution shortly after
saturation of the linear instability that was recently observed in
\cite{East2015}, an overview of which is provided in
\Section{linear-instability}. The difference between the two runs is visually
evident. While the three-dimensional solution becomes increasingly smooth at
late times, the two-dimensional one maintains a network of abrupt field
reversals. These structures are force-free rotational current layers, and are
examined in depth in \Section{current-layers}. As we will see in
\Section{energy-relaxed}, the total magnetic energy dissipated is dramatically
greater in the three-dimensional case than in the two-dimensional case. Both
runs show evidence of the inverse cascade; large-scale coherency of the magnetic
field must result from dynamical transfer of some magnetic energy toward longer
wavelengths since the initial spectrum is monochromatic around $k = \alpha_0$.
The inverse cascade will be examined in detail in \Section{inverse-cascade}.

\subsection{Linear instability of the excited Taylor states} \label{sec:linear-instability}

\figureEnergyDecayTimeSeries
\figureHelicityDistribution

Our helical initial conditions are linear force-free equilibria. Clearly they
are stable when $\alpha_0 = 1$ since such states are global energy minima for a
given magnetic helicity. The question of the ideal stability of the periodic
shorter wavelength ($\alpha_0 > 1$) Taylor states has a conflicted history
\cite[e.g.][]{Moffatt1986, Galloway1987, Er-Riani2014}, which has recently been
resolved in \cite{East2015}. In that study, numerical simulations of both FFE
and relativistic MHD revealed that $\alpha_0 > 1$ Taylor states are linearly
unstable to ideal perturbations. The instability is marked by exponential growth
of the electric field on roughly the \Alfven wave crossing time of the initial
structure size $\alpha_0^{-1}$, and saturates when the medium attains the
\Alfven speed, which for the magnetically dominated case is $c$, implying the
existence of regions where $E \approx B$. This instability affects generic
states, and the only counter-examples that were found were one-dimensional ABC
states (e.g.  $B_1=1,B_2=0,B_3=0$) that are stable for all values of
$\alpha_0$. Such states are pure plane waves having circular polarization, and
are force-free by virtue of having uniform magnetic pressure. All of our helical
initial conditions are short wavelength and either two or three dimensional, and
turbulent relaxation begins after the saturation of the ideal instability.

\subsection{Energy of fully relaxed configurations} \label{sec:energy-relaxed}

\figureCurrentSheetProfile
\figureAlphaDistribution

Here we discuss the magnetic energy associated with the end-state magnetic
configurations. Since the Taylor states have $\V B = \alpha_0 \V A$, their
energy is given simply by $\alpha_0 H/2$. In other words, their energy is
$\alpha_0$ times larger than the theoretical lower limit imposed by assuming the
state reaches $\alpha=1$ at constant $H$. Whether or not dynamical relaxation
processes settle with the field in this global energy minimum remains an open
question, but here we provide some evidence to support the following conjecture:
``force-free magnetic relaxation starting from periodic Taylor states ends in
the lowest energy configuration allowed by helicity conservation if and only if
the domain is three-dimensional.''

We have carried out a suite of calculations belonging to one of four categories,
being either two or three dimensional, and either helical or non-helical. Those
that are non-helical are, by construction, out of equilibrium at $t=0$ and could
decay until they reach zero energy since helicity conservation does not place
any lower bound on their magnetic energy. The helical ones are initially at an
unstable equilibrium, enter a period of turbulent relaxation, and settle in a
force-free equilibrium of lower energy. We considered initial conditions that
are both of the randomized type given by \Equation{general-ic} and the ABC type
given by \Equation{ABC}.

\figureAlphaHistCompare
\figurePowerSpectrumFit

Our results are summarized in \Figure{energy-decay-time-series}, which shows the
time series $U_B$ for each of six different runs. As expected, the non-helical
configurations decay toward zero energy in both two and three dimensional
settings. In three dimensions, the helical configurations all terminate in the
lowest energy state allowed by helicity conservation \footnote{The state of
  lowest allowed energy can be referred to \cite[e.g][]{Shats2005} as the
  spectral condensate.}. Both a randomized setup with $\alpha_0^2=257$ and the
2D ABC setup with $\alpha_0^2=256$ showed the same general behavior. The latter
setup was intentionally chosen to be identical with the two-dimensional setup
apart from the inclusion of a low-level (one part in $10^8$) white-noise
perturbation introduced to break the $z$-translational symmetry.

Both of the helical two-dimensional runs terminate their relaxation with an
energy that is decreased to only $30\%$ of its initial value. For example, a
randomized 2D initial condition with $\alpha_0 \approx 256$ settles in a state
that has roughly 77 times more magnetic energy than the Taylor minimum energy
state. This is not unique, as actually all of our helical 2D runs where
$\alpha_0 \gg 1$ (including $\alpha_0 = 16$) settle in a state whose energy is
decreased to roughly $30\%$. In \Appendix{numerical-convergence} we show that,
for the 2D ABC $\alpha_0 = 256$ setup, the final energy is numerically
converging to a value very near $30\%$. This means that the terminal states in
two dimensions are not linear equilibria. In other words, they do approximately
solve the force-free condition \Equation{force-free-condition}, but the value of
$\alpha$ may vary from one field line to another.

\subsection{Helicity distribution} \label{sec:hel-distrib}

The fact that the 2D configurations do not relax into linear equilibria stems
from invariance of the helicity distribution $\mathcal{H}(\psi)$ which we
introduced in \Section{topology}. \Figure{helicity-dist} confirms that it does
indeed remain constant over time to a very good approximation, even while the
volume distribution over the magnetic potential $d\mathcal{V}/d\psi$ changes
significantly. The feature evident in the bottom panel of
\Figure{helicity-dist}, where in the relaxed state most of the volume is
occupied by the extreme values of the magnetic potential, can be connected to
the formation of current layers which we discuss in the next section.

As an illustration that invariance of the helicity distribution might prevent 2D
relaxation from attaining longer wavelength linear equilibria, consider that the
magnetic potential function $A_z$ of any 2D linear equilibrium satisfies
\begin{equation*}
  A_z = \left(\frac{H}{2 \alpha_0 L^3} \right)^{1/2} \frac{B_z}{\bar{B}_z}
\end{equation*}
where $\bar{B}_z$ is the root mean square value of $B_z$ and $H$ is the total
helicity. $\mathcal{H}(\psi)$ can be characterized by its domain, namely the
global maximum of $|A_z|$ which we denote by $\psi_{\rm{max}}$. For two
different linear equilibria with frequencies $\alpha_0$ and $\alpha_0'$ to have
the same helicity distribution, it would be necessary for them to at least share
the same values of $H$ and $\psi_{\rm{max}}$, requiring that
\begin{equation} \label{eqn:alpha-ratio}
 \left( \frac{\alpha_0}{\alpha_0'} \right)^{1/2} = \frac{B_{z,\rm{max}} /
   \bar{B}_z}{B'_{z,\rm{max}} / \bar{B}'_z}.
\end{equation}
where $B_{z,\rm{max}}$ is the global maximum of $|B_z|$. Note that in general,
$\bar{B}_z \le B_{z,\rm{max}}$. For the particular case of the 2D ABC state $B_1
= B_2 = 1$, \Equation{alpha-ratio} leads to the requirement that $\alpha_0' \ge
\alpha_0 / 2$, and therefore its relaxation into a linear state with twice
smaller energy is impossible. In other words, there is no way for the 2D ABC
state represented in \Figure{helicity-dist}, whose $\alpha_0 = 32$, to evolve
into another linear equilibrium whose frequency is smaller than $16$, while
preserving $\mathcal{H}(\psi)$. We suspect this argument could be generalized
further, but for now we leave it as a conjecture that the wavelength of a linear
2D equilibrium may be uniquely specified by its helicity distribution --- which
if true would render it impossible for one linear 2D equilibrium to relax into
another of lower energy.

\subsection{The current layers} \label{sec:current-layers}

\figureMagneticBubbleProfile

During the turbulent relaxation of helical two-dimensional configurations, the
solution consists of oppositely signed flux domains separated from one another
by a network of rotational force-free current layers. The flux domains (black
and white regions of \Figure{a256-H1-3072-imgs}) have nearly uniform $B_z$ and
are thus relatively current-free. Across the current layers, the magnetic field
direction rotates through approximately $180^\circ$, while its magnitude (and
thus magnetic pressure) remains fixed. One such current layer is shown in
\Figure{a256-H1-3072-prof1d}, where a one-dimensional profile has been taken
along the $x$-axis, passing through the layer where it is aligned with the
$y$-axis.

It is evident that the current layers have a characteristic frequency, which we
denote by $\alpha_c$ and determine empirically as follows. Since the
solution is near a force-free equilibrium, the current is $\V J \approx \alpha
\V B$ for some spatially dependent frequency
\begin{equation*}
  \alpha = \V B \cdot \nabla \times \V B / B^2.
\end{equation*}
We anticipate that the probability density function $P(\alpha)$ will have two
local maxima --- one at $\alpha=0$ corresponding to the potential flux domain
interiors and the other at the frequency $\alpha_c$, marking the frequency of
the current layers. \Figure{a256-H1-alpha-hist} confirms this to be the case. It
shows $P(\alpha)$ at logarithmically spaced times throughout the relaxation, and
reveals the location of the second peak once the solution is sufficiently close
to a force-free equilibrium. For this particular run, with $\alpha_0 = 256$ and
$N = 3072$, the value of $\alpha_c \approx 128$.

It turns out to be a coincidence that in this case $\alpha_c \approx \alpha_0 /
2$. We have performed a family of calculations, varying $\alpha_0$ between 8 and
128, and varying $N$ between 128 and 8192. For each run, we recorded the value
of $\alpha_c$, time-averaged over roughly 100 snapshots between time $t=10$ and
$t=16$, which was late enough that the second peak in $P(\alpha)$ had emerged in
each run. There was no secular evolution of
$\alpha_c$. \Figure{alpha-hist-compare} reveals that current layers become
increasingly narrow with higher resolution, but \emph{also} with increasing
initial frequency $\alpha_0$. The scaling is consistent with the expression 
\begin{equation} \label{eqn:clayer-scaling}
  \alpha_c = k_1^{3/4} \alpha_0^{1/4}
\end{equation}
where $k_1 = N / 30$ is the turbulence cutoff frequency, which has been
determined by modeling the magnetic energy spectrum, (see
\Figure{a256-H1-16384-Pbfit}) and is insensitive to initial conditions,
depending only on the numerical scheme and grid resolution. We emphasize that
\Equation{clayer-scaling} is strictly empirical, in that it matches the data
shown in \Figure{alpha-hist-compare}.


The scaling given by \Equation{clayer-scaling} indicates that in the infinite
resolution limit, the current layers will have zero characteristic length.  We
are not able to say whether the scaling could be associated with a physical
property of 2D FFE at finite Reynolds number, or if it might depend on the
details of the numerical scheme. This question could be resolved by imposing the
turbulence cutoff frequency $k_1$ explicitly, and then varying the numerical
resolution. In other words, solutions to a resistive FFE system at a given
conductivity parameter will be necessary.

\subsection{Solitary magnetic bubbles}

\figurePspecEvolve
\figureScaleEvolution

The flux domain interiors contain another type of coherent structure. They are
long-lived bubbles of toroidal magnetic field that are confined by ambient
magnetic pressure. The circular object to the left of the current layer in
\Figure{a256-H1-3072-prof1d} is an example. They may be referred to as flux
tubes or magnetic vortices, but we will refer to them as magnetic bubbles since
we believe they are similar objects to those studied by \cite{Gruzinov2010}. The
bubbles are helical, force-free magnetic field structures, having $\V J$ very
well aligned with $\V B$. But they are not linear equilibria; the value of
$\alpha$ varies with distance $r$ from the axis, and all the bubbles we examined
had similar radial profiles $\alpha(r)$. The current flow is parallel to the
background magnetic field near the core, but an equal and opposite return
current flows through the sheath. So they could also be thought of as co-axial
electric current channels oriented along the symmetry axis.

Force-free configurations with axial and $z$-symmetry satisfy the ordinary
differential equations
\begin{eqnarray} \label{eqn:cyl-force-free}
  B_\phi \alpha + B'_z = 0 \\
  B_\phi r^{-1} - B_z \alpha + B'_\phi = 0 \nonumber
\end{eqnarray}
where $\alpha(r)$ needs to be specified to make the solution unique, as well as
the boundary conditions $B_z(r) \rightarrow B_\infty$ and $B_\phi(r) \rightarrow
0$ as $r \rightarrow \infty$. Requiring the total current $I = \int B_\phi r d
\phi$ to be zero only forces $B_\phi(r)$ to decrease faster than
$1/r$. Alternatively, the magnetic pressure $u_B(r)$ may be specified instead of
$\alpha(r)$. One simple assumption, consistent with measurements of the bubbles,
is that the magnetic pressure enhancement, relative to the ambient pressure
$B^2_\infty/2$, is Gaussian. This gives a pressure profile
\begin{equation} \label{eqn:bubble-ub}
  u_B(\tr) = \frac{1}{2} B^2_\infty (1 + f e^{-\tr^2})
\end{equation}
in the dimensionless radius $\tr = \alpha_b r$ for a bubble of radius
$\alpha_b^{-1}$, where $f$ is the magnetic pressure enhancement relative to
background and could be any positive number. The corresponding solution of
\Equation{cyl-force-free} is
\begin{eqnarray*}
  B_\phi(\tr) &=& B_\infty f^{1/2} \tr e^{-\tr^2/2} \\
  B_z(\tr) &=& B_\infty \sqrt{1 + f e^{-\tr^2} (1 - \tr^2)} \\
\end{eqnarray*}
which has the profile
\begin{equation} \label{eqn:bubble-alpha}
  \alpha(\tr) = \alpha_b \frac{(2 - \tr^2) e^{-\tr^2/2}}{\sqrt{(1 - \tr^2)
      e^{-\tr^2} + f^{-1}}}.
\end{equation}

In \Figure{magnetic-bubble-profile} we show the radial profile of magnetic
pressure, azimuthally averaged around the bubble's axis. This is the same object
depicted to the left of the current layer in \Figure{a256-H1-3072-prof1d}. We
also show the expression given in \Equation{bubble-ub} with its best-fit model
parameters. For this particular object, the best-fit fractional magnetic energy
enhancement was $f = 0.41$ and the best-fit inverse radius was $\alpha_b =
162$. The lower panel of \Figure{magnetic-bubble-profile} shows the radial
profile of $\alpha$, and also the model predicted value given by
\Equation{bubble-alpha} with the best-fit parameters.

So far we have not detected these structures in 3D simulations, but that does
not mean they never happen. It is crucial to address whether they are even
stable in three dimensions, and if they are, then under what conditions they may
be attractors.

\subsection{The inverse cascade} \label{sec:inverse-cascade}

The inverse cascade can be characterized by the rate with which the magnetic
coherence scale $k_B^{-1}$ (\Equation{k_X}) migrates toward longer wavelengths.
We observe inverse cascading in force-free electrodynamics for every setting we
have considered, whether the turbulence is 2D or 3D, helical or
non-helical. \Figure{pspec-evolve} shows evolution of the magnetic energy
spectrum $P_B(k)$ over time, for representative helical and non-helical runs in
2D. We note how in both cases, the remaining magnetic energy resides at an
increasing scale over time. We also note the spectral energy density at long
wavelengths to be an increasing function of time, indicating that
\emph{selective decay} of the short wavelength modes alone cannot explain growth
of the spatial coherency. This further generalizes observations made of
non-helical 3D MHD turbulence, in both Newtonian \citep{Brandenburg2015} and
relativistic \citep{Zrake2014} settings.

\Figure{scale-evolution} shows the evolution of the characteristic magnetic
frequency $k_B$ for one model from each of the four categories. For the helical
runs we also show the evolution of the helicity frequency $k_H$. Except for the
general feature that over time, both $k_B$ and $k_H$ move to smaller frequency,
there is no single decay law that characterizes all these settings. Some runs
exhibit power-law time dependence of $U_B$, $k_B$, or $k_H$, but not all.
Power-law dependence is considered evidence of self-similarity in the relaxation
process, meaning the solution evolves only by rescaling itself in space and time
\citep[e.g.][]{Landau1987}. Self-similar evolution may occur when the
characteristic scales are all sufficiently smaller than the domain size, so that
processes around the coherence scale are not yet contaminated by the requirement
of periodicity at the domain scale. For this reason, large values of $\alpha_0$,
and thus high domain resolution, may be required for self-similarity to
emerge. For 3D, freely decaying non-helical relativistic MHD turbulence,
\cite{Zrake2014} found power-law dependence of $k_B \propto t^{-2/5}$ for
$\alpha_0 \approx 48$ initial conditions. For the same conditions in FFE, $k_B$
evolves with a power-law index of $-0.48$, as shown in the second column of
\Figure{scale-evolution}. However, the dependence of $k_B$ on time is not
convincingly power-law, indicating that a scale separation of $\alpha_0 \approx
48$ may not be sufficient to yield a decisive measurement of the decay index. In
the future, we plan to follow-up on this with higher resolution simulations.

In freely decaying 2D non-helical force-free turbulence with very high
resolution ($12288^2$) and $\alpha_0 \approx 256$, clear power-law behavior of
$k_B$ is observed, with an index of $-0.38$. The helical 2D case is
different. As shown in \Figure{scale-evolution} (column 3), the energy drops to
its terminal value of $\approx 0.3 U_{B}(t=0)$ long before $k_B^{-1}$ approaches
the domain scale. At times later than $t=0.15$, the merging of flux domains
(evident in \Figure{a256-H1-3072-imgs}) moves magnetic energy to progressively
larger scales while suffering slower and slower dissipative losses. During this
epoch, the both the helicity and magnetic frequencies decay with a power-law
index of roughly $-0.55$.

In general, freely decaying magnetic turbulence must exhibit an inverse cascade
when the magnetic helicity is near maximal ($H \approx U_{B} / k_B$), if
$\partial_t H = 0$ is to be satisfied \citep{Frisch1975, Christensson2001,
  Cho2011}. However, the same conclusion cannot be drawn from magnetic helicity
conservation alone when $H \ll U_{B} / k_B$, as is the case for our non-helical
runs. Observation of the non-helical inverse cascade was thus seen as a surprise
\citep{Zrake2014, Brandenburg2015}.  It was suggested in \cite{Zrake2014} that
the non-helical inverse cascade may reflect the tendency for aligned current
structures to attract one another. \cite{Brandenburg2015} found that net
transfer of energy from small to large scales was about twice larger in helical
than non-helical freely decaying Newtonian MHD turbulence.


\subsection{Power spectrum of electric and magnetic energy} \label{sec:pspec}

\figurePspecCompare

In all of our initial conditions the magnetic energy is concentrated around a
single frequency, $P_B(k) \propto \delta(k - \alpha_0)$. As turbulence develops,
energy redistributes itself over all available scales, with the bulk of the
energy around $k_B$ (which decreases over time as discussed
in \Section{inverse-cascade}) and a power-law tail extending to the spectral
cutoff frequency $k_1$ which lies consistently near $N/30$ when the grid
resolution is $N$. We have determined the index $s$ of the power-law tail by
fitting the logarithm of the spectral distributions $P_E(k)$ and $P_B(k)$ to the
model function
\begin{equation} \label{eqn:spectral-fit}
  f(x) = A e^{-(x - x_0)^2 / 2\sigma^2} + x s
\end{equation}
where $x = \log k$, and $A$, $x_0$, $\sigma$, and $s$ are model parameters. We
have found it expedient not to try and model the high-frequency cutoffs, we just
use frequency bins between the peak frequency $\tilde{k}_B$ and the cutoff $k_1$
to obtain our fit. A representative spectral fit is shown in
\Figure{a256-H1-16384-Pbfit}.

\Figure{pspec-compare} shows $P_E(k)$ and $P_B(k)$ for one model from each of
the categories 2D/3D and helical/non-helical along with the best-fit model given
by \Equation{spectral-fit}. We use randomized initial conditions for each model,
and the resolution is $16384^2$ in 2D and $1024^3$ in 3D (coinciding dash-dotted
curves are taken from simulations whose resolution is $12288^2$ and $768^3$ to
indicate numerical consistency). The spectra represent the solution at an
intermediate time, when $k_B \approx 8$, and we find the power-law indices to be
quite stable \footnote{Between $t=0.6$ and $t=1.2$, the spectral indices show
  negligible secular evolution, and a standard deviation with respect to
  different time levels that is below the level of $1\%$. The spectral indices
  reported are the instantaneous values, which are within a standard deviation of
  the mean.} during a window of time beginning on the snapshot chosen for each
model. The power-law index of the electric field spectral energy distribution
$P_E(k)$ is between 0.96 and 1.18 for the various models, with both of the
helical models having an index of 1.18. The power-law index of the magnetic
field spectral energy distribution $P_B(k)$ is different in two and three
dimensions. In both helical and non-helical 3D settings, it is notably close to
the Kolmogorov value of 5/3. Note that \cite{Zrake2014} and
\cite{Brandenburg2015} both measured an index near 2 for freely decaying
non-helical MHD turbulence. The helical 2D model has an index of 1.93, and it is
tempting to conclude that the true value is 2, but actually the best-fit
parameter, for all resolutions up to $16384^2$ is significantly smaller than 2,
always being between 1.92 and 1.96. The non-helical 2D model is found to have an
index of 1.41.

\section{Discussion and conclusions} \label{sec:discussion}

Using high resolution simulations of the force-free electrodynamics equations,
we have studied freely decaying, magnetically dominated relativistic turbulence
for the first time. We focused on various differences between the magnetic
relaxation process in settings where the domain is two and three dimensional,
and where the magnetic field is helical and non-helical. We found that helical,
two-dimensional relaxation terminates in a state whose energy is far above the
theoretical minimum imposed by helicity conservation, and whose volume is
predominantly current-free, but is punctuated by coherent structures --- namely
current layers and solitary magnetic bubbles. We tried to determine what sets
the width of the current layers, and determined that it depends not only on the
turbulence cutoff frequency (or grid resolution), but also on the frequency
$\alpha_0$ of the initial condition. The solitary magnetic bubbles are
axisymmetric, non-linear force-free equilibria, and are consistent with Gaussian
magnetic pressure enhancement relative to the surrounding relatively
current-free volume. The three-dimensional stability of these structures remains
an open question.

The unusual behavior of 2D relaxations can be understood in terms of additional
topological constraints that are imposed by the extra symmetry. We proposed that
unlike generic 3D relaxation, whose topology is only constrained by the total
helicity invariant $H$, 2D relaxations are subject to a whole spectrum of
helicity invariants $\mathcal{H}(\psi)$, one associated with each value of the
magnetic potential $\psi$. Although this invariance is only guaranteed when
magnetic reconnections are restricted to regions of zero volume, our simulation
data showed that $\mathcal{H}(\psi)$ remains unchanged throughout the evolution
to a very good approximation.

All of the settings we considered exhibited inverse cascading, in which some
magnetic energy is redistributed toward progressively longer wavelengths. The
rate of inverse cascading was characterized by the time evolution of the
magnetic frequency $k_B$, which was found to decrease faster when the field is
helical than non-helical, and also faster in 3D than in 2D. The inverse cascade
of 2D helical turbulence is nearly conservative; merging of magnetic flux
domains moves energy to larger scales while suffering a diminishing rate of
dissipative losses. The non-helical inverse cascade has $k_B \propto t^{-0.38}$
in 2D and $k_B \propto t^{-0.48}$ in 3D, in rough agreement with results of
\cite{Zrake2014} for 3D turbulent relaxation in relativistic MHD. Better scale
separation (larger $\alpha_0$) and thus higher numerical resolution is needed to
confirm the three dimensional scaling measurement.

\subsection{Astrophysical gamma-ray sources and magnetoluminescence}
\label{sec:astro-implications}

We have examined turbulent relaxation in force-free electrodynamics with the
motivation 
of elucidating the physical mechanism behind extremely fast time variability
that is characteristic of astrophysical gamma-ray emitters, including the Crab
Nebula, many blazars, and nearly all gamma-ray bursts. The extreme energetics
and temporal intermittency of gamma radiation from these sources require a
mechanism in which plasma promptly converts the majority of its magnetic energy
into high energy particles and radiation. Furthermore, the emitting regions are
thought to be strongly magnetized, and known to lie a great distance from the
primary mover (pulsar, progenitor star, or black hole). These facts are highly
suggestive that electromagnetic outflows may contain persistent magnetic
structures with copious free energy supplies, whose spontaneous disruption could
be linked to the observed flaring events. A scenario like this was referred to
in \cite{Blandford2014} as \emph{magnetoluminescence}. For it to be plausible,
it is necessary that (1) meta-stable, force-free (or hydromagnetic) equilibria
can exist far from any supporting boundaries, (2) that such objects can form
under realistic astrophysical conditions, and (3) that upon their disruption,
magnetic energy is promptly and completely dissipated. 


In this paper, we have begun to address the points (1) and (3) and found results
that are at least partially encouraging. All of our periodic 3D simulations
exhibit prompt relaxation into the Taylor minimum energy state, supporting the
idea that magnetic energy can be dissipated completely in a light-travel time.
But the same result suggests that persistent, meta-stable structures are not a
generic outcome of turbulence in force-free electrodynamics. This is not to say
that such behavior is impossible, as we have only considered a small class of
initial conditions. In fact, \cite{Smiet2015} very recently identified magnetic
arrangements that on 3D periodic domains, in full MHD, relax to non-linear
hydromagnetic equilibria. So (1) is possible, at least in the hydromagnetic
case. The volatility of such objects, and the generality of conditions under
which they may arise remain important questions for the future.


\subsection{Comparison with other studies of magnetic relaxation}

In this study we have begun to address the question of whether relativistic,
force-free magnetic relaxation on periodic domains generically ends in a Taylor
state, and found evidence to support the view that it does, provided the domain
is three dimensional. But thus far we have only considered a restricted class of
initial conditions --- namely isotropic, monochromatic fields that are either
linear force-free equilibria (helical) or completely non-helical.

Several studies in full MHD have now identified settings that relax to more
general \emph{hydromagnetic} equilibria for which $\V J \times \V B = \nabla p$,
where the Lorentz force density balances the gradient of gas pressure
$p$. Examples include stratified three-dimensional environments
\citep{Braithwaite2006, Braithwaite2008} where the field is helical, and
two-dimensional periodic settings where the field is incompressible and
non-helical \citep{Gruzinov2009}. \cite{Amari2000} and \cite{Braithwaite2015}
have both provided examples of 3D hydromagnetic relaxation which first develop
current layers, and then proceed to a smooth configuration via resistive
processes. Both studies used boundary conditions where at least one of the
directions was not periodic. Very recently, \cite{Smiet2015} has found instances
of 3D hydromagnetic relaxation ending in smooth, non-linear equilibria with
non-uniform pressure, even when the boundary is periodic or open. Such boundary
conditions are highly relevant for the astrophysical processes mentioned
in \Section{astro-implications}, and an important question is whether their
results possess any force-free analogues. In particular, if stable and
non-linear force-free equilibria do exist in 3D away from boundaries, then how
likely are they to arise under realistic astrophysical conditions?

\subsection{Effects of imposing extra symmetries in astrophysical simulations}

Many studies of relativistic plasma and MHD processes are carried out assuming
either translational or rotational symmetry, including simulations of force-free
electrodynamics \citep{McKinney2006, Tchekhovskoy2008}, relativistic MHD
\citep{Barkov2008, Komissarov2009, Komissarov2009a, Mizuno2011}, and
particle-in-cell (PIC) simulations \citep{Spitkovsky2008, Keshet2009}. As we
have seen, 2D magnetic relaxations are far more topologically constrained, and
persist in configurations of much higher energy than equivalent 3D relaxations.
Furthermore, axisymmetric calculations are expected to exhibit similar
artificialities to the slab-symmetric ones studied here, since they both share
the same topological simplifications. In particular, the axisymmetric magnetic
surfaces, now toroidal shells that are labeled by their value of the azimuthal
vector potential component $A_\phi$, each enclose a conserved magnetic helicity.

Our results indicate that as a field's complexity increases, so does the
discrepancy between the energy of its most relaxed state in 2D and 3D. So,
axisymmetry may be appropriate when the field is near a stable force-free
equilibrium, when little or no energy resides in higher-order radial or angular
modes. But when these modes \emph{are} populated, the imposed symmetry could
make them artificially persistent.

Numerical studies of pulsar magnetospheres and magnetar flares are quite
challenging, and much of the progress in this field has been obtained by
restricting to axisymmetry.  Simulations using both FFE \citep{Parfrey2012,
  Yu2012, Parfrey2013} and PIC \citep{Cerutti2015} show the development of
complex structure in the meridional plane.  Similarly, short wavelength
toroidal magnetic structure has been found in axisymmetric FFE simulations of
black hole accretion.  For example the results of \cite{Parfrey2014} suggest
that even disordered (as opposed to large-scale) magnetic fields advected inward
by an accretion disk could facilitate angular momentum extraction from a black
hole.  The differences in 2D versus 3D found here suggest that extending these
studies to be fully three dimensional, as that becomes computationally feasible,
will be an exciting frontier, and will likely reveal qualitatively new dynamics.


Another setting where 2D and 3D calculations are expected to differ is
shock-generated turbulence, which may account for the relatively high
magnetizations inferred from non-thermal emission spectra of astrophysical shock
fronts --- both in non-relativistic settings such as supernova remnants and
relativistic settings such as gamma-ray burst afterglows. Amplification of the
magnetic field by turbulence in and around the shock, and its subsequent decay
in the post-shock flow has been studied extensively in both two
\citep{Mizuno2011} and three \citep{Inoue2010} dimensions with relativistic MHD,
and also in 2D and 3D PIC simulations \citep{Sironi2009, Sironi2013a}. In this
study we have observed power-law decay of the magnetic energy in all settings,
but with a steeper index in 3D than in 2D. This should be kept in mind, as even
minor differences in the decay law can have an impact on the efficiency of first
order Fermi processes \citep{Lemoine2006, Niemiec2006, Niemiec2006a,
  Pelletier2009} as well as interpretations of GRB afterglows
\citep{Gruzinov1999a, Rossi2003, Lemoine2014}.

\acknowledgments The authors are grateful for extensive discussions with Yajie
Yuan, Krzysztof Nalewajko, and Roger Blandford, and also for the continued
guidance and encouragement of Tom Abel, Andrew MacFadyen, and Andrei Gruzinov.
We also thank Luis Lehner for helpful comments. Simulations were run on the
Bullet Cluster at SLAC and the Sherlock Cluster at Stanford University, and also
on Comet at the San Diego Supercomputer Center (SDSC) through XSEDE grant
AST150038, as well as Pleiades of the NASA High-End Computing (HEC) Program
through the NASA Advanced Supercomputing (NAS) Division at Ames Research Center.

\newpage
\appendix

\section{Numerical convergence} \label{sec:numerical-convergence}

\figureTseriesConvergence

Here we demonstrate some numerical convergence properties of our scheme. We have
chosen the conservation of magnetic helicity $\Delta H(t)$ and the time series
of magnetic energy $U_B(t)$ as diagnostics. Convergence properties are reported
for the 2D helical runs with $\alpha_0 = 256$ and grid resolutions $4096^2$,
$6144^2$, $8192^2$, $12288^2$ and $16384^2$. This configuration was found to be
representative of convergence properties in other settings reported in this
paper. All runs were evolved for at least two light-crossing times. In order to
establish the numerical convergence order $n$, we model the error of the
numerical solution $y_h$ with grid spacing $h$ as $y_h = y_0 + E h^n$ where $E$
is a constant and $y_0$ is the extrapolated solution. This is similar to a
Richardson extrapolation, but instead of fitting for the coefficient $E_n$ for
each integer power of $h$, we have fit for the single error coefficient $E$ and
convergence order $n$.

The upper left panel of \Figure{a256-H1-tseries-convergence} shows the
fractional change in magnetic helicity $H(t) / H_0 - 1$ as a function of time
for each resolution. For resolutions $>4096^2$, the helicity change is never
worse than $\pm 10\%$. For $>8192^2$ it is never worse than $\pm 1\%$, and for
$>12288^2$ it is never worse than $\pm 0.1\%$. The extrapolated value of the
helicity change is consistently a gain of about $0.1\%$, and the convergence
order (shown on the lower left panel of \Figure{a256-H1-tseries-convergence}) is
between 2.8 and 2.9. The right panel shows the evolution of magnetic energy
$U_B(t)$ at each resolution. Less dissipation occurs for each higher resolution,
but the sequence converges consistently at first order. The extrapolated value
of the magnetic energy at $t=2$ is 0.30, and it changes by less than $1\%$
between $t=1$ and $t=2$.


\bibliographystyle{apj}
\bibliography{ref,library}

\end{document}

%% file: figures.tex
\newcommand{\figureImageSequence} {
  \begin{figure*}
    \centering
    \includegraphics[width=6.8in]{\figext{a256-H1-3072-imgs}}
    \includegraphics[width=6.8in]{\figext{a16-H1-512-3d-imgs}}
    \caption{\textbf{Top}: Two-dimensional turbulent relaxation in force-free
      electrodynamics at logarithmically spaced times
      ($t=0.08,0.32,1.28,5.12$). The initial condition is the $\alpha_0=256$ ABC
      field with $B_1=1,B_2=1,B_3=0$ and grid resolution $3072^2$. Shown is the
      out-of-plane magnetic field component scaled linearly between the initial
      minimum and maximum values. The small red rectangle overlying the
      right-most panel is the region shown amplified in
      \Figure{a256-H1-3072-prof1d}. The end-state is not a linear force-free
      equilibrium. \textbf{Bottom}: Three-dimensional turbulent relaxation under
      the same conditions except that $\alpha_0=16$, the grid resolution is
      $512^3$, and the times $t=0.625,1.0,3.0,16.0$ are chosen to elucidate the
      sequence of decay epochs. The color mapping accomodates the instantaneous
      data range, as it decreases appreciably throughout the decay. The
      end-state is a linear force-free equilibrium with $\alpha=1$.}
    \label{fig:a256-H1-3072-imgs}
  \end{figure*}
}

\newcommand{\figureEnergyDecayTimeSeries} {
  \begin{figure}
    \centering
    \includegraphics[width=3.6in]{\figext{energy-decay-time-series}}
    \caption{Total magnetic energy $U_B$ as a function of time (since the onset
      of nonlinear evolution $t_0$). The energy of the terminal state takes on
      one of three values. In the case of two-dimensional helical evolution, the
      end-state contains coherent structures and retains $30\%$ of its initial
      energy, whereas for three-dimensional evolution the end-state is a longest
      wavelength Taylor state whose energy is reduced by a factor of $\alpha_0$
      from its initial energy. When the decay is non-helical magnetic energy
      decays perpetually toward zero. The randomized initial condition $\alpha_0
      \approx 16$ corresponds to $\alpha_0^2 = 257$ and $\alpha_0 \approx 256$
      corresponds to $\alpha_0^2 = 65641$.}
    \label{fig:energy-decay-time-series}
  \end{figure}
}

\newcommand{\figureHelicityDistribution} {
  \begin{figure}
    \centering
    \includegraphics[width=3.6in]{\figext{helicity-dist}}
    \caption{\textbf{Top} --- The helicity distribution $d\mathcal{H}/d\psi$
      versus the level surface value $\psi$ of the magnetic potential function,
      shown at 16 evenly spaced times up to $t=16$ in a 2D ABC run with
      $\alpha_0=32$ and resolution $4096^2$. As anticipated in
      \Section{topology}, $d\mathcal{H}/d\psi$ is essentially constant in time.
      \textbf{Middle} --- The volume distribution $d\mathcal{V}/d\psi$,
      indicating the volume between level surfaces at a given $\psi$. The
      separatrix surfaces $\psi=0$ initially occupy the greatest volume, become
      the current layers, and end up with the smallest share of the
      volume. \textbf{Bottom} --- The helicity per volume
      $d\mathcal{H}/d\mathcal{V}$. Lighter, wider curves indicate earlier time
      whereas darker, thinner curves indicate later times. The distributions are
      each normalized to unity.}
    \label{fig:helicity-dist}
  \end{figure}
}

\newcommand{\figureCurrentSheetProfile} {
  \begin{figure*}
    \centering
    \includegraphics[width=7.2in]{\figext{a256-H1-3072-prof1d}}
    \caption{Amplification of the small red rectangle overlying the rightmost
      panel of \Figure{a256-H1-3072-imgs}, showing relief plots of the $x$, $y$,
      and $z$ components of the magnetic field (on the top three panels). The
      bottom panel shows one-dimensional profiles, taken along the horizontal
      centerline (dashed magenta).}
    \label{fig:a256-H1-3072-prof1d}
  \end{figure*}
}

\newcommand{\figureAlphaDistribution} {
  \begin{figure}
    \centering
    \includegraphics[width=3.6in]{\figext{a256-H1-alpha-hist}}
    \caption{Probability density function $P(\alpha)$ at logarithmically spaced
    times. The right-most vertical dashed line is the frequency $\alpha_0=256$
    of the initial field configuration. The local maximum at $\alpha \approx
    128$ is the frequency $\alpha_c$ of the current sheets (which is resolution
    dependent), and the maximum at $\alpha=0$ corresponds to zero-current
    characterizing the flux domain interiors.}
    \label{fig:a256-H1-alpha-hist}
  \end{figure}
}

\newcommand{\figureAlphaHistCompare} {
  \begin{figure}
    \centering
    \includegraphics[width=3.6in]{\figext{alpha-hist-compare}}
    \caption{Depenency of the current layer frequency $\alpha_c$ on the
      frequency $\alpha_0$ of the initial field configuration (top) and the grid
      resolution $N$ (bottom). $\alpha_c$ is defined to be the second local
      maximum (other than $\alpha = 0$) in the probability density function
      $P(\alpha)$, where $\alpha = \V B \cdot \nabla \times \V B / B^2$.}
    \label{fig:alpha-hist-compare}
  \end{figure}
}

\newcommand{\figurePowerSpectrumFit} {
  \begin{figure}
    \centering
    \includegraphics[width=3.6in]{\figext{a256-H1-16384-Pbfit}}
    \caption{Example best-fit (shown in red) to magnetic power spectrum data
      ($+$'s) for a two-dimensional helical model with $\alpha_0=256$ and
      resolution $16384^2$. $P_B(k)$ is shown compensated by the best-fit
      spectral index $s=1.94$ (top) and uncompensated (bottom). Vertical dashed
      lines indicate the window used for the fit, between the peak magnetic
      frequency $\tilde{k}_B$ and the spectral cutoff frequency $k_1$.}
    \label{fig:a256-H1-16384-Pbfit}
  \end{figure}
}

\newcommand{\figureMagneticBubbleProfile} {
  \begin{figure}
    \centering
    \includegraphics[width=3.6in]{\figext{magnetic-bubble-profile}}
    \caption{\textbf{Top} --- The magnetic pressure profile of a magnetic bubble
      in the dimensionless cylindrical radius $\tilde{r} = \alpha_b r$. The blue
      shaded region indicates the azimuthal standard deviation at each radius
      from the center, and the dashed line shows the best-fit model parameters
      for the Gaussian mangetic pressure enhancement given by
      \Equation{bubble-ub}. \textbf{Bottom} --- The azimuthally averaged value
      of $\alpha$ along with its predicted value (dashed line) given by
      \Equation{bubble-alpha}. The top and bottom insets show two-dimensional
      relief plots of $u_B$ and $\alpha$, respectively.}
    \label{fig:magnetic-bubble-profile}
  \end{figure}
}

\newcommand{\figurePspecEvolve} {
  \begin{figure}
    \centering
    \includegraphics[width=3.6in]{\figext{pspec-evolve}}
    \caption{Magnetic energy spectra $P_B(k)$ shown at logarithmically spaced
      times for a 2D simulation of freely decaying force-free
      turbulence. Randomized initial conditions with $\alpha_0 \approx 256$ and
      $N = 16384$, where $\epsilon=0$ (non-helical) on top and $\epsilon=1$
      (helical) on bottom.}
    \label{fig:pspec-evolve}
  \end{figure}
}

\newcommand{\figureScaleEvolution} {
  \begin{figure*}
    \centering
    \includegraphics[width=6.8in]{\figext{scale-evolution}}
    \caption{Total magnetic energy $U_B$ as a function of time (top row) and
      evolution of the magnetic frequency $k_B$ and, for the helical runs, the
      helicity frequency $k_H$ (bottom row). One representative run is chosen
      from each of the four settings --- 2D/3D non-helical, and 2D/3D
      helical.}
    \label{fig:scale-evolution}
  \end{figure*}
}

\newcommand{\figurePspecCompare} {
  \begin{figure}
    \centering
    \includegraphics[width=3.6in]{\figext{pspec-compare}}
    \caption{Power spectra of the electric (top) and magnetic (bottom) energy
      densities as given by \Equation{power-spectra} for one model from each of
      the catagories 2D/3D and helical/non-helical. The solid curves are
      measured from simulations of whose resolution is $16384^2$ in 2D and
      $1024^3$ in 3D, while the coinciding dash-dotted curves are measured from
      simulations whose resolution is slightly lower, $12288^2$ and
      $768^3$. Dashed lines indicate the best-fit parameters of
      \Equation{spectral-fit}. An example fit is shown in more detail in
      \Figure{a256-H1-16384-Pbfit}.}
    \label{fig:pspec-compare}
  \end{figure}
}

\newcommand{\figureTseriesConvergence} {
  \begin{figure*}
    \centering
    \includegraphics[width=7.2in]{\figext{a256-H1-tseries-convergence}}
    \caption{\textbf{Left} - Error in the conservation of magnetic helicity
      $H$. The upper panel shows the fractional helicity change $\Delta h(t) =
      H(t)/H_0 - 1$ on symmetric logarithmic axes (to account for anomolous
      helicity change of either sign) for six different values of the mesh spacing
      $h$. The dashed magneta line shows the Richardson-extrapolated value of
      $\Delta h(t)$, which remains constant at roughly $10^{-3}$. The lower panel
      shows the convergence order of $\Delta h(t)$ at representative
      times. \textbf{Right} - Evolution of the total magnetic energy $U_B(t)$ for
      the same six values of the mesh spacing. The dashed magenta line on the
      upper panel shows the Richardson-extrapolated time series of $U_B(t)$, and
      the convergence order is shown on the lower panel.}
    \label{fig:a256-H1-tseries-convergence}
  \end{figure*}
}